\documentclass[a4paper,fleqn,usenatbib]{mnras}

\usepackage{multicol}
\usepackage[T1]{fontenc}
\usepackage{ae,aecompl}

\usepackage{graphicx}	
\usepackage{amsmath}	
\usepackage{amssymb}	

\title[TT Arietis: 40 years of photometry]
{TT Arietis: 40 years of photometry}

\author[A. Bruch]{Albert Bruch
\\
Laborat\'orio Nacional de Astrof\'{\i}sica, Rua Estados Unidos, 154, 
CEP 37500-364, Itajub\'a, MG, Brazil
}

\date{Accepted XXX. Received YYY; in original form ZZZ}

\pubyear{2019}

\begin{document}
\label{firstpage}
\pagerange{\pageref{firstpage}--\pageref{lastpage}}
\maketitle

\begin{abstract}
In an effort to characterize variations on the time scale of hours and 
smaller during the high and low states of the novalike variable TT~Ari, light 
curves taken over the course of more than 40~yr are analyzed. It is found 
that the well known negative superhump observed during the high state 
persists until the present day at an average period of 0.13295~d 
which is slightly variable from year to year and exhibits substantial 
amplitude changes. The beat period between superhump and orbital period 
is also seen. QPOs occur at a preferred quasi-period of 18 -- 25~min and 
undergo a systematic frequency evolution during a night. The available 
data permit for the first time a detailed investigation of the low state 
which is highly structured on time-scales of tens of days. On hourly time 
scales the light curve exhibits strong variations which are mostly irregular. 
However, during an interval of several days at the start of the low state, 
coherent 1.2~mag oscillations with a period of 8.90~h are seen. During the 
deep low state quiet phases and strong (1.5 -- 3~mag), highly structured 
flares alternate in irregular intervals of roughly 1 day. The quiet phases 
are modulated on the orbital period of TT~Ari, suggesting reflection 
of the light of the primary component off the secondary. This is the first 
time that the orbital period is seen in photometric data.
\end{abstract}

\begin{keywords}
stars: activity -- {\it (stars:)} binaries: close -- 
{\it (stars:)} novae, cataclysmic variables -- stars: individual: TT Ari 
\end{keywords}



\section{Introduction}
\label{Introduction}

TT~Ari (= BD+14$^{\rm o}$341), detected as a variable star by 
\citet{Strohmeier57},
is one of the brightest and best studied members of the novalike subclass of
cataclysmic variables (CVs). As such, it consists of a late type dwarf star
(the secondary) that transfers matter via Roche lobe overflow to a white dwarf
(the primary), forming an accretion disk around the latter. For a comprehensive
overview of all aspects of CVs and their various subtypes, see 
\citet{Warner95}. 

Due to a high mass transfer rate from the secondary, the accretion disk in 
novalike variables remains in a high brightness state, disabling 
instabilities which give rise to outbursts such as those observed in dwarf
novae \citep{Lasota01}. Thus, in general, their long term light curve does not 
contain strong variations. However, in some systems, named after their 
prototype VY~Scl, the mass transfer rate sometimes
is reduced to much lower values and the luminosity of the accretion disk falls
by up to several magnitudes. The occurrence and the duration of these low
states is unpredictable. TT~Ari belongs to the VY~Scl stars. Drops from the
high state at $V \approx 10.8$~mag to a low state reaching down to
$V \approx 16.3$~mag (noting that just as in all VY~Scl stars the low
state brightness does not remain constant but is highly variable) have
been observed in 1979 -- 1985 and 2009 -- 2011. \citet{Wu02} derive
component masses of $M_1 = 1.24\ M_\odot$, $M_2 = 0.23\ M_\odot$ and an orbital
inclination of $i = 29^{\rm o}$, but warn that these values may be uncertain. The
distance, based on the Second Gaia Data Release, is $256 \pm 5$~pc
\citep{Bailer-Jones18}.

Regular variations with a period of about 3.2~h were first observed during 
the high state by \citet{Smak69}. These
were initially considered to be orbital in nature. However,
spectroscopic observations by \citet{Cowley75}, later confirmed by
\citet{Thorstensen85}, revealed the true orbital period to be slightly
longer. The currently most accurate value is  
$P_{\rm orb} = 0.13755040 \pm 1.7 \times 10^{-7}$~d  
\citep[3.3012086~h;][]{Wu02}. Nowadays, the photometric modulation is
interpreted as a negative superhump, caused by the variation of the depth
within the gravitational potential of the white dwarf of the impact point
of the transferred matter onto an accretion disk which is inclined with 
respect to the orbital plane and therefore suffers nodal precession in the 
fixed reference frame of the binary star \citep{Wood07}. The superhump 
period is then the beat period between the precession and the orbital periods.
This negative superhump of TT~Ari has been studied many times in the past
\citep{Smak69,Sztajno79,Semeniuk87,Roessiger88,Udalski88,Volpi88,Andronov92,
Andronov99,Tremko92,Kraicheva99,Kim09,Weingrill09,Belova13}

While the negative superhump has been present during most of the observational
history of TT~Ari when it was in the high state, in the time interval between
1997 and 2004 it was replaced by another modulation at a period slightly 
longer than the orbital period. First detected by \citet{Skillman98},
and later also characterized by \citet{Kraicheva99}, \citet{Wu02}
and \citet{Belova13}, it is interpreted as a positive superhump. This
phenomenon, commonly seen in SU~UMa type dwarf novae during superoutbursts,
is thought to be caused by the apsidal precession of an elliptically 
deformed accretion disk \citep{Whitehurst88, Hirose90}. Theory predicts 
that such ellipticity can only be induced into the disk if its radius 
attains the one third resonance radius with the orbit of the secondary 
star. This would restrict the occurrence of elliptical disks to short orbital 
period systems with a low secondary-to-primary mass ratio 
\citep{Whitehurst91}. It is therefore an unsolved 
question why positive superhumps are also occasionally observed in systems 
with a much higher mass ratio and a wider orbit. Prominent examples are, 
among others, 
UU~Aqr \citep{Patterson05},
V603~Aql \citep[][and references therein]{Bruch18},
KR~Aur \citep{Kozhevnikov07},
TV~Col \citep{Retter03},
V751~Cyg \citep{Patterson01, Papadaki09},
V795~Her \citep{Patterson94, Papadaki06}, and
V378~Peg \citep{Kozhevnikov12}.
TT~Ari another case.

Apart from superhumps, strong flickering, ubiquitous in CVs, modulates the 
light of TT~Ari. Not well separated from flickering are flaring events
which occur on the time-scale of roughly 20~min and are termed quasi-periodic
oscillations (QPOs) in the literature. Along with the superhumps they have 
also been subject to many of the studies cited above.

Compared to the high state, the low states of TT~Ari are much less 
investigated. The 1979 -- 1985 low state was covered by \citet{Shafter85}, 
\citet{Hutchings85}, and \citet{Gaensicke99}. The only
detailed paper dedicated to the 2009 -- 2011 low state was published by
\citet{Melikian10}. The low state photometry of \citet{Shafter85}
reveals flickering, indicating residual mass transfer even
during the lowest observed brightness, while \citet{Melikian10} report
strong flares superposed on a baseline level of about the same magnitude.

In the present paper I analyse extensive archival photometric observations 
spanning more than 40 yr, in order to further characterize variations on
various time-scales. In Sect.~\ref{The data} the data are briefly
presented. Subsequently, in Sect.~\ref{The high state} the high state data
are investigated with emphasis on the negative superhumps, noting that from
now on, the term 'superhump', unless explicitly stated otherwise, always 
is meant to refer to the {\it negative} superhump. QPOs and flickering are
also being looked at in some details. The 2009 -- 2011 low state is
subject of Sect.~\ref{The 2009 -- 2011 low state}, investigating the rich and
so far untapped observational data collected in particular by the American
Association of Variable Star Observers (AAVSO). Conclusions are summarized in
Sect.~\ref{Conclusions}.

\section{The data}
\label{The data}

This study is based on data originating from a variety of sources.
The bulk was retrieved from the AAVSO International Database which contains
the long term light curve of TT~Ari since 1977. While until the year 2000 
almost all AAVSO data consist of visual magnitudes
which are not used here, in later years a growing number of observations were 
performed in (or reduced to the) the $V$ band and, to a lesser degree, the
$B$, $R$ and $I$ bands. A few observations performed in the $U$ band are not 
considered. Many of the $B$, $V$ and $R$ band observations represent time
resolved light curves, i.e., continuous data sets spanning several hours at
a time resolution between 10 and 120~s.

To these data I add light curves taken from miscellaneous sources. Apart
from the publically available $R$ band data first investigated by 
\citet{Kim09}, I use light curves kindly provided by a variety of
observers. These include unpublished white light data from E.\ Nather,
$B$ and $U$ band data from I.\ Semeniuk [partly discussed in 
\citet{Semeniuk87} and \citet{Schwarzenberg-Czerny88}], S.\ R\"o{\ss}iger 
\citep[$B$ band; see][]{Roessiger87}, A.\ Hollander \citep[light curves in the 
Walraven system, discussed in][]{Hollander92}, unpublished $UBV$ data 
taken by T.\ Schimpke, and light curves observed in the Stiening photometric 
system \citep{Horne85}, provided by E.\ Robinson. The latter have an
extremely high time resolution of 0.5~s and were binned here into intervals
of 5~s in order to reduce noise.

The entire data set comprises more than 160\,000 individual data points. 
This is arguably the largest ensemble of observations of TT~Ari ever
investigated.

The time stamps of all data were reduced to Barycentric Julian Date on the
Barycentric Dynamical Time scale, using the online tool of 
\citet{Eastman10}. This was not necessary for the \citet{Kim09} light
curves because these are already expressed in Heliocentric Julian Date (thus
neglecting the small difference between barycentric and heliocentric time). 

\begin{figure}
	\includegraphics[width=\columnwidth]{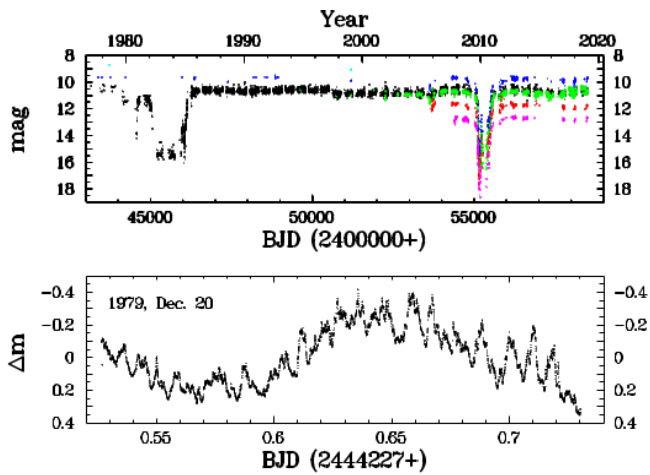}
    \caption{{\it Top:} The long term light curve of TT~Ari binned in 1~d
                 intervals. The black dots represent visual magnitudes.
                 $U$, $B$, $V$, $R$ and $I$ measurements are shown in
                 cyan, blue, green, red and magenta, respectively. For clarity,
                 the $B$ and $U$ band data are shifted upwards by 1~mag and 
                 and $R$ and $I$ band data downwards by 1 and 2~mag, 
                 respectively.
                 {\it Bottom:} Time resolved light curve of 1979 December 20
                 as a typical example, showing clear superhump variations 
                 together with flickering and QPOs.}
    \label{overall-lc}
\end{figure}

Fig.~\ref{overall-lc} (top) contains the overall long term light curve. Visual 
magnitudes are shown as black dots. $I$, $R$, $V$, $B$ and (the few) $U$ data 
are colour coded in magenta, red, green, blue and cyan, respectively. For 
clarity, the $B$ and $U$ band data are shifted upwards by 1~mag, and the $R$
and $I$ band data downwards by 1 and 2~mag, respectively. All magnitudes have 
been binned in
1 day intervals. Since the data provided by E.\ Nather, I.\ Semeniuk, 
and E.\ Robinson are not calibrated, an average magnitude is assigned to them.
As an example of the typical variations of TT~Ari on the time-scale of hours,
the lower frame of Fig.~\ref{overall-lc} shows the (unfiltered) light curve of 
1979 December 20. The variations caused by the 3.2 h superhump are outstanding.
Superposed is random flickering and (not clearly separated from flickering) a 
flaring activity with a limited regularity, identified as QPOs. 

During most of the more than 40 yr of observations TT~Ari remained in an
almost stable (except for some small glitches) state of high brightness. 
However, as is well documented in the literature, during two time intervals 
it went into a deep low state, first for a prolonged period between the end 
of 1979 and the beginning of 1985, and then
again for a shorter time between September 2009 and February 2011.
The first of these was covered only by visual observations with a coarse
time resolution. It is therefore not regarded here. The second low state,
however, was extensively observed in $B$, $V$, $R$ and $I$, many times at a high
time resolution. These observations reveal some interesting new features
of TT~Ari (see Sect.~\ref{The 2009 -- 2011 low state}).

\section{The high state}
\label{The high state}

\subsection{Superhumps}
\label{Superhumps}

As mentioned in Sect.~\ref{Introduction}, TT~Ari is well known to exhibit
negative superhumps
during most of the time it spends in the high state. Only during the brief
interval between 1997 -- 2004 they were replaced by positive superhumps.
While they have been documented several times in the literature, the 
present data set permits a more rigorous assessment of its properties 
and evolution than has been possible in the past.

\subsubsection{The extensive AAVSO data of 2012, 2014 and 2017}
\label{The extensive AAVSO data of 2012, 2014 and 2017}

\begin{figure}
	\includegraphics[width=\columnwidth]{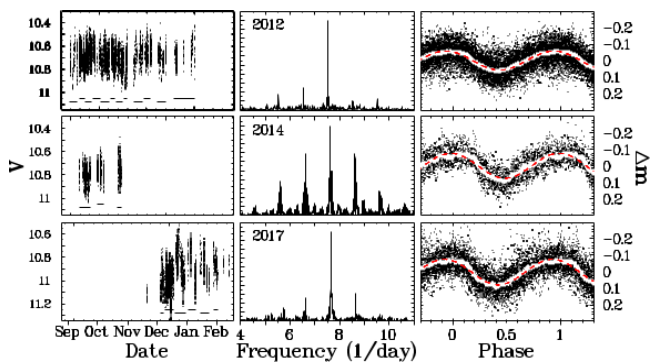}
      \caption[]{{\it Left column:} Combined $V$ band light curves observed 
                 during the 2012 (top), 2014 (middle) and 2017 (bottom)
                 observing seasons. The bars beneath 
                 the light curves indicate the intervals used to search
                 for possible variations of the superhump period (see text).
                 {\it Central column:} Power spectra of the combined light
                 curves. {\it Right column:} Light curves after subtraction 
                 of the night-to-night variations, folded on the period 
                 corresponding to the dominant peak in the power spectra. 
                 The white ribbons represent the same light curves, binned
                 in phase intervals of width 0.01. The red broken graph is a
                 least squares sine fit to the binned light curves.}
\label{2012-2017-sh}
\end{figure}

By far the most high state data, best suited for the purpose of this
study, refer to the years after the second low state. Most of them consist 
of time resolved light curves, spanning up to several hours. For a timing 
analysis of the superhumps in these data
I restrict myself to the $V$ band, and then only to the
time resolved light curves having a resolution of better than 100~s
and spanning at least 1~h. The most extensive data refer to the 2012
observing season\footnote{The observing
season of TT~Ari normally extends from about August of year $n$ to 
February -- March of year $n+1$. To facilitate notation, I will always
refer myself to the observing season of year $n$, even if some observations
were obtained at the beginning of the following year.}, encompassing a total 
of 112 light curves (see Fig.~\ref{2012-2017-sh}, upper left frame). Most 
of the scatter in magnitude seen during individual nights is due to regular
variations which can readily be identified as being caused by the
superhump. Superposed, slight variations on longer time-scales are seen.
These appear not to be random, indicating that differences in the calibration
of individual light curves, which may be expected considering that the data
were obtained by a variety of observers, remain small.  

In a first step the nightly mean was subtracted from the individual light
curves. Then a Lomb-Scargle periodogram
\citep[][hereafter also referred to as power spectrum]{Lomb76, Scargle82} 
of the combined data set was calculated.
A sharp peak dominating the spectrum provided the preliminary period of the 
superhump variations which are clearly seen in the individual light curves. 
In the next step a sine curve of the form 
$V = a \sin \left( 2\pi (t-t_o)/P \right) + \gamma$ was fit to each light 
curve. Here, $a$ is the half amplitude, $t$ the time of the observations,
$t_o$ the zero point of phase, $P$ the period fixed to the preliminary value, 
and $\gamma$ is an offset in magnitude. This offset was then subtracted from 
each light curve. This procedure provides a better subtraction of 
night-to-night variations
than simply subtracting the mean. The resulting combined
data set was again submitted to the Lomb-Scargle algorithm, and the resulting
power spectrum is shown in the upper central frame of Fig.~\ref{2012-2017-sh}. 

The power spectrum is dominated by a strong peak, accompanied by secondary 
peaks at both sides which are 1~d$^{-1}$ aliases. The maximum provides the 
superhump period $P_{\rm SH} = 0.1328743$ d, listed in the second column of 
Table~\ref{superhump period} together with the error that (here and in all 
similar cases later) is propagated from the error in
frequency, conservatively defined as the standard deviation of a 
Gaussian fit to the maximum in the power spectrum. The light curve, cleaned 
from night-to-night variations and folded on this period, is shown in the
right upper frame of Fig.~\ref{2012-2017-sh}, where 
the maximum of the superhump waveform is taken to be the 
zero point of phase. The white ribbon in the folded light curve 
represents a binned version, 
using intervals of 0.01 in phase. The red broken line is a least squares
sine fit, yielding a full amplitude $A_{\rm sh}$ of the 
modulation\footnote{Hereafter, whenever reference to the amplitude of
variations in a light curve is made, the full difference between minium
and maximum is meant.} listed in the last but one column of 
Table~\ref{superhump period}.

\begin{table*}
	\centering
	\caption{Properties of negative superhumps and related structure
	in the TT~Ari light curves of observing seasons 2012, 2014 and 2017.
        Periods are expressed in days, amplitudes in magnitudes. The difference
        between $P_{\rm long}$ and $P_{\rm prec}$ is expressed in units of its
        statistical error.}
	\label{superhump period}

\begin{tabular}{lllllll}
\hline
year & \phantom{$\pm$}$P_{\rm SH}$ & \phantom{$\pm$}$P_{\rm long}$ & 
\phantom{$\pm$}$P_{\rm prec}$ & $P_{\rm long} - P_{\rm prec}$ &
\phantom{$\pm$}$A_{\rm SH}$   & \phantom{$\pm$}$A_{\rm long}$ \\
\hline
2012 & \phantom{$\pm$}0.132874 &
       \phantom{$\pm$}3.98    &
       \phantom{$\pm$}3.90858 &
                    --1.91      &
       \phantom{$\pm$}0.116       &
       \phantom{$\pm$}0.065       \\
     &           $\pm$0.000037 &
                 $\pm$0.02     &
                 $\pm$0.03208 &
                                &
                 $\pm$0.003 &
                 $\pm$0.006 \\ [1ex]
2014 & \phantom{$\pm$}0.132799  &
       \phantom{$\pm$}3.83    &
       \phantom{$\pm$}3.84456   &
                    --0.08   &
       \phantom{$\pm$}0.152       &
       \phantom{$\pm$}0.074       \\
     &           $\pm$0.000139 &
                 $\pm$0.17  &
                 $\pm$0.11295 &
                               &
                 $\pm$0.011 &
                 $\pm$0.008 \\ [1ex]
2017 & \phantom{$\pm$}0.132643  &
       \phantom{$\pm$}{\it 4.12}    &
       \phantom{$\pm$}3.71806   &
       \phantom{$\pm$}{\it 3.50}   &
       \phantom{$\pm$}{\it 0.154}       &
                                \\
     &           $\pm$0.000092  &
                 {\it $\pm$0.09}   &
                 $\pm$0.40344 &
                              &
                 {\it $\pm$0.001} & \\
\hline
\end{tabular}
\end{table*}

A close inspection of the folded light curve reveals that the superhump
wave form is not quite sinusoidal. The decline from maximum is slightly
steeper than the ascent. This asymmetry should be reflected in the power
spectrum as signals at harmonic frequencies of the main peak. Indeed,
it contains a weak maximum at $2 f_{\rm SH} = 15.052$~d$^{-1}$.

Next, the stability of the period was investigated. For this purpose, 12
sections of the whole 2012 light curve (marked by horizontal lines below
the light curve in Fig.~\ref{2012-2017-sh}) were subjected to the 
Lomb-Scargle analysis. These sections are identified in 
Table~\ref{period stability} which also lists the superhump periods. 
The forth column contains the difference between the average period 
during the observing season and the period measured in the specific 
light curve section, expressed in units of the $1\sigma$ uncertainty of 
that difference. In only one section (BJD 2456227 -- 2456230) no 
consistent periodic variations were obvious and in one other section 
(BJD 2456213 -- 2456217) the peak frequency is significantly different from 
the average of the entire data set. All other subsections exhibit peak 
frequencies comfortably consistent within their error margins with the 
overall frequency. Therefore, there is no 
evidence for variations of the superhump period over the entire $\approx$4 
month spanning the observations. The last column of the table contains
the amplitudes of the modulations in the respective sections. They exhibit
erratic variations which are significantly larger than their formal errors,
these being of the order of 2 mmag.

\begin{table}
	\centering
	\caption{Periods (in days) and amplitudes (in magnitudes) of negative 
        superhumps in restricted sections of the TT~Ari light curves of 
        observing seasons 2012, 2014 and 2017. The last but one column 
        contains the difference between the period within a section and the
        average seasonal period, expressed in units of its statistical error.}
	\label{period stability}

\begin{tabular}{lllll}
\hline
Year & BJD      & Period & Diff.      & Amplitude \\
\hline
2012 & 2456169 -- 177 & 0.132857 &           --0.03  & 0.156 \\
     & 2456179 -- 185 & 0.132700 &           --0.26  & 0.131 \\
     & 2456185 -- 192 & 0.132758 &           --0.19  & 0.131 \\
     & 2456193 -- 202 & 0.133184 & \phantom{--}0.46  & 0.110 \\
     & 2456202 -- 211 & 0.132431 &           --0.64  & 0.119 \\
     & 2456213 -- 217 & 0.133921 & \phantom{--}1.61  & 0.101 \\
     & 2456219 -- 226 & 0.133035 & \phantom{--}0.28  & 0.136 \\
     & 2456227 -- 230 & \multicolumn{1}{r}{--}  & \multicolumn{1}{r}{--} &
                        \multicolumn{1}{l}{\phantom{0.00}--} \\
     & 2456236 -- 246 & 0.132797 &           --0.14  & 0.085 \\
     & 2456248 -- 255 & 0.132691 &           --0.31  & 0.120 \\
     & 2456262 -- 271 & 0.132848 &           --0.04  & 0.122 \\
     & 2456279 -- 301 & 0.132800 &           --0.17  & 0.126 \\ [1ex]
2014 & 2456909 -- 921 & 0.132862 &           --0.02  & 0.160 \\
     & 2456928 -- 935 & 0.131581 &           --1.67  & 0.106 \\
     & 2456949 -- 953 & 0.132857 &           --0.01  & 0.195 \\ [1ex]
2017 & 2458090 -- 094 & 0.132380 &           --0.20  & 0.163 \\ 
     & 2458095 -- 105 & 0.132475 &           --0.41  & 0.173 \\
     & 2458106 -- 117 & 0.133037 & \phantom{--}0.22  & 0.140 \\
     & 2458119 -- 128 & 0.132505 &           --0.50  & 0.195 \\
     & 2458132 -- 142 & 0.132709 &           --0.23  & 0.174 \\
     & 2458146 -- 151 & 0.133145 & \phantom{--}0.21  & 0.149 \\
\hline
\end{tabular}
\end{table}

\begin{figure}
	\includegraphics[width=\columnwidth]{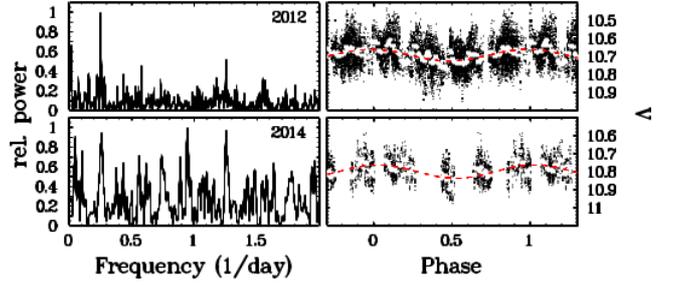}
      \caption[]{{\it Left:} Power spectra of the combined 2012 (top) and
                 2014 (bottom) light curves after removal of the superhump 
                 modulations. {\it Right}, combined lines curves, folded on
                 the period corresponding to the power spectrum peak close to
                 0.25~d$^{-1}$. The white dots represent the same data,
                 binned in phase intervals of width 0.01. The red curves are
                 least squares sine fits to the binned light curves.}
\label{beat}
\end{figure}

In order to search for other periodic signals in the light curve, the best
fitting sine curve with a period fixed to the superhump period was subtracted 
from each individual light curve and then the offset $\gamma$ of the fitted 
sine was added again. In this way the superhump variations are removed, but the
combined light curve still contains variations on longer (and shorter) time 
scales. The corresponding power spectrum is shown in the upper left frame of
Fig.~\ref{beat}. As listed in the third column of 
Table~\ref{superhump period}, the strongest
peak corresponds to a period of $P_{\rm long} = 3.9808$~d. On the other hand,
the orbital period together with $P_{\rm SH}$ yield a beat period, interpreted
as the precession period of the accretion disk, of $P_{\rm prec} = 3.90857$~d
(column 4 of Table~\ref{superhump period}). Thus, the difference 
$P_{\rm long} - P_{\rm prec} = -1.91 \sigma$ (column 5 of 
Table~\ref{superhump period}), where $\sigma$ is the error of 
$P_{\rm long} - P_{\rm prec}$. Although $P_{\rm long}$ is not quite identical to 
$P_{\rm prec}$, 
considering the formal error limits, it appears that the precession of the 
accretion disk induces a slight variability in the brightness of the system,
confirming earlier results of \citet{Semeniuk87}, \citet{Udalski88}
and \citet{Kraicheva97}. The light curve folded on $P_{\rm long}$ is
shown in the upper right frame of Fig.~\ref{beat}. The white dots
are a binned version of the folded light curves, using intervals of width 
0.01 in phase. The red solid line is a least squares sine fit to the binned
light curve. The amplitude is listed in the last column of 
Table~\ref{superhump period}.

Similar exercises were performed for the 2014 and 2017 observing seasons, 
using 16 light curves observed between 2014 September 8 and October 22, and 
48 light curve obtained between 2017 November 18 and 2018 February 12. The 
results are shown in Fig.~\ref{2012-2017-sh} and
listed in Tables~\ref{superhump period} and \ref{period stability}. The 
superhump amplitude is similar in both seasons, but higher than observed
in 2012. The waveform of the superhump deviates from a pure sine in the
same way as in 2012. No significant period changes occur during the
smaller observing intervals, but in 2017 the average period is
slightly smaller than in previous seasons.

The power spectra of the residual light curves, after removing
the superhump variation, do not reveal the beat frequency between orbit 
and superhump as clearly as in 2012. In 2014 (lower frames of 
Fig.~\ref{beat}) two peaks at different frequencies (not obviously related to 
other periodicities in the system) are about as strong as the maximum at 
0.2612~d$^{-1}$
which I take to correspond to $P_{\rm long}$. However the close coincidence 
between $P_{\rm long}$ and $P_{\rm prec}$ and the amplitude of the corresponding 
variations which is quite similar to that measured in 2012 lend some 
confidence that a brightness modulation on the precession period really
exists. 

In 2017 this issue is less clear. The power spectrum of the residual 
light curve is rather inconclusive because the average magnitude
of TT~Ari varied much more during this season than in previous years (see
Fig.~\ref{2012-2017-sh}). It is
unclear whether these night-to-night variations are real or caused by
uncertainties of the magnitude calibration of the light curves compiled from
different sources. The power spectrum has only a minor peak close to the 
expected precession frequency. The corresponding period is quoted 
tentatively in Table~\ref{superhump period} (in italics, in order to 
distinguish it from other, more reliable entries in the table),
but the difference $P_{\rm long} -
P_{\rm prec}$ being 3.5 times the error margin casts doubt on the
presence of a modulation on the disk precession period. 
It is not sensible to quantify the amplitude of the light curve folded on
that period in view of the much stronger night-to-night variations.

\subsubsection{Re-analysis of the Kim et al.~(2009) data}
\label{Re-analysis of the Kim et al. (2009) data}

\citet{Kim09} observed TT~Ari in the $R$ band during 48 nights between
2005 October 27 and 2006 March 6. These data cover the recovery of the
system from a minor downward glitch in its magnitude 
(see Fig.~\ref{overall-lc}). TT~Ari rose from an average magnitude of 
$R = 11.25$ to a local brightness maximum of $R = 10.3$ and then settled 
down at an intermediate level of about $R = 10.7$ 
\citep[see fig.~1 of ][]{Kim09}. \citet{Kim09} used times
of minima in the light curves to measure the superhump period, but did
not comment on the fact that superhumps were not present during the entire
period of their observations.

\begin{figure}
	\includegraphics[width=\columnwidth]{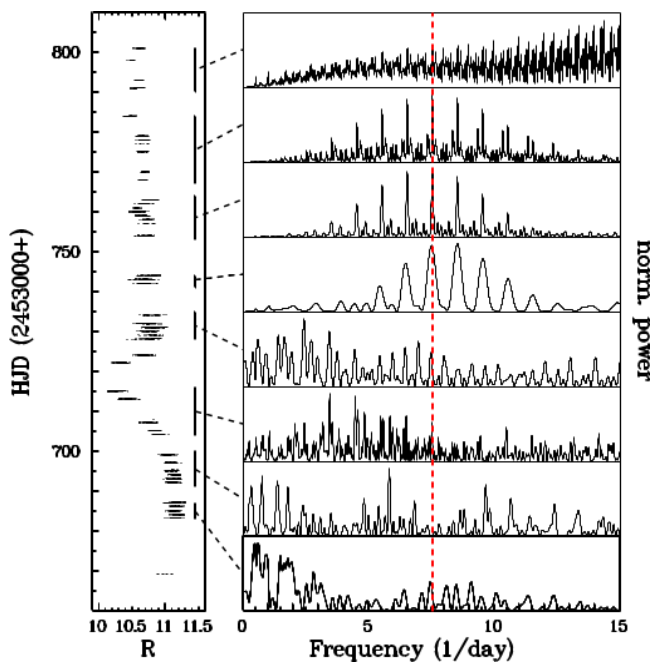}
      \caption[]{The combined light curves observed by \citet{Kim09}
                 (left) is shown together with power spectra (right) calculated 
                 from the light curve sections marked by vertical bars. The
                 broken vertical line marks the frequency of the superhump 
                 modulation derived from the combined light curves in the 
                 interval HJD~2453741 -- 84).}
\label{Kim powerspectra}
\end{figure}

A re-analysis of their data, following the same procedures outlined in
Sect.~\ref{The extensive AAVSO data of 2012, 2014 and 2017},
reveals the evolution of superhumps over time. This is shown in 
Fig.~\ref{Kim powerspectra}. On the left hand side, the light curve is
repeated. Vertical bars indicate intervals referring to the power
spectra drawn on the right hand side of the diagram. These clearly show
that a periodic signal at the superhump frequency (the broken vertical line,
derived from the power spectrum of the combined data in the interval
HJD~2453741 -- 84) is at most marginally present during the faint state
of TT~Ari until just after the local maximum. Thereafter, the superhump
evidently dominates the light curves. Somewhat surprisingly, during the
end of the entire interval covered by \citet{Kim09}, it may still be
present but is no longer outstanding.
The development of the fully grown superhump occurs remarkably fast.
Some trace may be present in the interval HJD~2453714 -- 35, but only 6~d
later (HJD 2453741) it is the dominant feature in the power spectrum.
These time-scales should provide limits for models explaining their formation 
and destruction.

The superhump period of $0.132624 \pm 0.000150$~d, derived from the light 
curve of the time interval when the superhump is clearly present, is slightly 
longer than that quoted by \citet{Kim09} who used a different method for
its determination. The amplitude is 0.094~mag. When comparing it to the
amplitude observed at other epochs it should be remembered that it
refers to $R$ band observations. As observed in most of the AAVSO data,
the superhump waveform is slightly asymmetrical with a slower rise to
maximum and a faster decline to minimum. No convincing evidence for a 
brightness modulation on the beat period between orbit and superhump was
found.
   
\subsubsection{Smaller data sets observed between 1977 and 2004}
\label{Smaller data sets observed between 1977 and 2004}
 
\begin{figure}
	\includegraphics[width=\columnwidth]{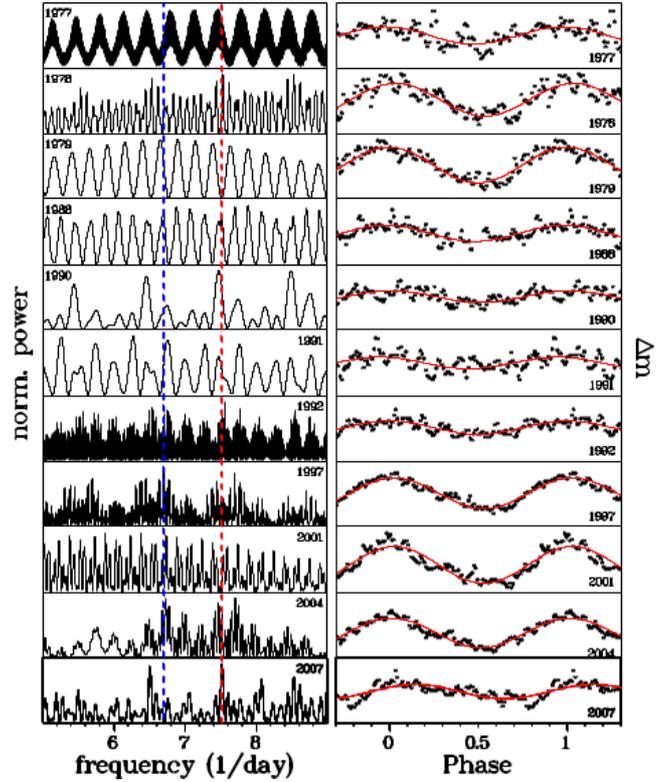}
      \caption[]{{\it Left:} Power spectra of fragmentary light curves
                 observed in various years. The red (right) vertical line 
                 indicates the long-term average frequency of the negative 
                 superhump, while the blue (left) 
                 vertical line is the positive superhump frequency.
                 {\it Right:} Light curves of the
                 respective years, after subtraction of the nightly mean,
                 folded on the period listed in 
                 Table~\ref{long term period stability} and binned in
                 intervals of 0.01 in phase. The vertical scale of each
                 diagram encompasses 0.3~mag, except for 1979 where
                 it is twice this value. The red curves are least
                 squares sine fits to the data.}
\label{ps-various-years}
\end{figure}

Smaller data sets are available for several other observing seasons. They
are briefly discussed here. In most but not all cases the power spectra 
contain more or less clear evidence for superhumps. While the corresponding
peaks are not always the strongest, but aliases, the closeness to the well
established period seen at other epoch lends confidence in their reality.
The resulting periods are summarized in Table~\ref{long term period stability} 
which also contains results compiled from the literature 
(see Sect.~\ref{The period stability}). It
also lists the amplitude of the modulation as determined from a least 
squares sine fit, and the passband to which the amplitude refers. 
Depending on the distribution of the individual light curves within the 
observing window cycle counts are sometimes uncertain and
thus the correct choice among the alias peaks in the periodogram
is not always unique. Therefore, errors based on the width of the peaks 
will be misleading and are not quoted in the table. 
The left hand side of Fig.~\ref{ps-various-years} shows the  power spectra for 
each investigated season while the right hand side contains the phase folded 
light curves, binned in intervals of width 0.01 and a least square sine fit.
The red (right) vertical line indicates the long-term average frequency of the
negative superhump modulation (see Sect.~\ref{The period stability}), 
while the blue (left) vertical line is the positive
superhump frequency as measured by \citet{Skillman98}.

In no case an assessment of the presence of variations on the beat period
between superhump and orbit is possible due to the small number of nightly
light curves and because many of them are nor calibrated and therefore lack
a common magnitude scale.

1977: Three unfiltered ('white light') light curves provided by E.\ Nather,
observed between 1977 August 9 and December 15 yield a power spectrum
with a peak (although not quite the highest) corresponding to a period of 
0.134049~d, close to but still somewhat longer than the average
superhump period.

1978: Two $U$ band and two $B$ band light curves observed between August 2 and 
25 were kindly provided by I.\ Semeniuk. The $U$ band data are also listed by
\citet{Schwarzenberg-Czerny88}. The power spectrum has a significant
peak quite close to the expected superhump frequency. Folding the light curves
on the respective periods indicates that the satellite peaks on both sides 
of the main one can be
discarded as representing the true period. The folded light curve shown
in Fig.~\ref{ps-various-years} and the amplitude quoted in 
Table~\ref{long term period stability} only refer to the 
$U$ band where the superhump is more clearly expressed than in $B$. 

1979: Only two light curves observed in white light, kindly provided by 
E.\ Nather,
are available. However, obtained within four days on 1979 December 20 and 24,
they are of extremely high quality and exhibit regular variations very clearly.
Due to the insufficient time coverage the power spectrum of the combined light 
curves contains a forest of alias peaks which makes it impossible to
select the ``correct'' one without pre-knowledge. The highest peak leads to 
a period of 0.150347~d, far from the expected negative superhump period, but
remarkably close to the positive superhump period observed 2~yr later. 
However, there are no other reports indicating that the negative has been
replaced by a positive superhump during this observing season. Only about
two weeks earlier, \citet{Semeniuk87} observed a negative superhump, albeit
at a period somewhat higher than the average (see 
Sect.~\ref{The period stability}). The currently discussed light curve has
a peak, only slightly lower than the highest one, which corresponds to 
$P = 0.130953$~d, similar to the period identified by \citet{Semeniuk87} and
which is listed in Table~\ref{long term period stability}. The amplitude 
of the modulations is at 0.33~mag much higher than observed at other epochs.

1985: Six $B$ band light curves provided by I.\ Semeniuk and S.\ R\"ossiger are
available. However, although in some of the individual light curves 
systematic variations on time-scales roughly comparable to the expected
superhump period are present, the overall light curve does not lead to
a consistent picture. Therefore, these data are not considered further here.

1988: Three Walraven light curves observed between 1988 August 16 and 27
were kindly provided by A.\ Hollander. They have been discussed by
\citet{Hollander92}. The power spectrum is dominated by a 
multitude of alias peaks. The maximum at 7.5152 d$^{-1}$, which I adopt here
as the most relevant, is not the strongest 
one, but the corresponding period (0.13306~d) is very close to the expected 
negative superhump period. The closest and slightly stronger alias peak 
corresponds to 0.1299~d, much smaller than ever observed for the superhump. 
Moreover, the light curve folded on the adopted period supports the 
presence of consistent variations on this period. Its amplitude, listed in
Table~\ref{long term period stability} refers to the $B$ band and was 
transformed from flux units as provided by the Walraven photometer 
\citep{Walraven60, Rijf69} into magnitudes.

1990: Six light curves observed in the Stiening system, kindly provided by E.\
Robinson are available. Five of them were observed between 1990 October
12 and 18. The last one is not considered here because of the long time
difference of 64~d to the others which complicates the power spectrum
significantly. The most prominent peaks in the power 
spectrum correspond to a period of 0.133831~d (rather far away from the
superhump period observed in other years) and to its 1~d$^{-1}$ alias. The 
folded light curve, even after binning, is quite noisy and has a low
amplitude, making it difficult to assess if the periodic modulation is 
spurious or real. Keeping this caveat in mind, the results are listed in 
Table~\ref{long term period stability}

1991: The situation in 1991 is quite similar to that in the previous year. 
Again, six Stiening light curves are available, with 5 of them observed in the 
time window between 1991 October 4 and 12, and the last one in 1991 December.
The latter is disregarded for the same reason as above. The power spectrum 
does not provide a clear picture. The highest peak corresponds to a period of 
0.159425~d, far away from any expected superhump period. One of its
aliases corresponds to 0.137334~d, more compatible but still
significantly different from the superhump period in other seasons. Again,
the folded light curve is of low amplitude and noisy, and is thus not quite 
convincing. Therefore, the same caveat expressed earlier must be kept in mind.

1992: Two data sets observed in 1992 are available. 
The first one consists of five $UBV$ light curves, kindly provided by
T.\ Schimpke, observed between 1992 August 10 and 17. 
Four light curves in the Stiening system, provided by E.\ Robinson,
obtained between September 24 and November 29 constitute the second set.
The power spectrum is complex. However, the strongest peak corresponds
to a period very close to the expected superhump period.

1997: By 1997, the negative superhump in TT~Ari gave way to a positive 
superhump \citep{Skillman98}. This is confirmed by five AAVSO light curves
observed between 1997 December 1 and 1998 January 22. This time window
is part of the larger window covered by \citet{Skillman98}. It is
therefore not surprising that the dominant
power spectrum peak yields a period of 0.14931 day, i.e., within the error
limit of the period quoted by \citet{Skillman98}, based on more extensive data.

2001: The power spectrum of three light curves observed between 2001 December 4 
and 24 has its strongest peak at a frequency corresponding to a period of
0.170048~d. However, there is an alias peak corresponding to 0.148441~d.
While this is significantly different from the period observed in 1997,
it is well known that in superoutburst of SU~UMa type dwarf novae the
(positive) superhump period is not stable but evolves over time 
\citep[see, e.g.,][]{Kato09}. The same may be expected here.
Moreover, between 2000 August and 2001 January, 
\citet{Stanishev01} observed a superhump period of 0.148815~d in the 
light curve as well as in the equivalent width of various emission
lines. Therefore, the modulation in the present data seen at 0.148441~d may
confidently be identified as the positive superhump in TT~Ari. Its 
amplitude is similar to that observed in 1997. 

2004: The power spectrum of six light curves observed between 2004 
December 4 and 25 shows that the positive superhump still persists 
\citep[see also][]{Andronov05}. 
The amplitude appears to be slightly smaller than in previous years.  

2007: Eight light curves are available for this season. They were observed in a 
comparatively small time window between 2007 October 23 and November 6. The 
superhump modulation is clearly present. The waveform deviates 
more strongly from a pure sine curve than in 
other seasons and in the opposite sense. It has a steeper rise to maximum 
and a more gradual decrease.

\subsubsection{The long term properties of superhumps}
\label{The period stability}

The period of the negative superhumps has been measured many times at
many different epochs. Therefore, its stability and systematic or
stochastic variations can be assessed. In their table 3, \citet{Tremko96}
present a list of period determinations up to 1988. Since then,
further measurements have become available which warrant an update, 
presented here in Table~\ref{long term period stability} together with 
the results of the present study. The table also includes (in italics) 
periods of the positive superhump observed between 1997 and 2004. A plot 
of the negative superhump period vs.\ time (not shown)
does not suggest a systematic variation over more than
half a century covered by data, but rather some small scale scatter. 
Averaged over the total time base the period is 
$0.13295 \pm 0.00067$ days.

\begin{table}
	\centering
	\caption{Compilation of superhump periods and their amplitudes
        of TT~Ari measured in different observing seasons since 1961} 
        \label{long term period stability}

\begin{tabular}{lllcl}
\hline
Year & Period & Amp.      & Pass-   & Reference \\
     & (days) & (mag)     & band    &           \\
\hline

1961 & 0.1329   &       &       & \citet{Smak75}               \\
1966 & 0.1327   &       &       & \citet{Smak75}               \\
1977 & 0.1338   &       &       & \citet{Semeniuk87}           \\
1977 & 0.13405  & 0.079 & white & this work                    \\
1978 & 0.13280  & 0.176 & $U$   & this work                    \\
1978 & 0.1326   & 0.2   & $B$   & \citet{Sztajno79}            \\
1979 & 0.13095  & 0.327 & white & this work                    \\
1985 & 0.132771 &       &       & \citet{Roessiger88}          \\
1985 & 0.132765 &       &       & \citet{Udalski88}            \\
1985 & 0.1328   &       &       & \citet{Semeniuk87}           \\
1986 & 0.13298  &       &       & \citet{Volpi88}              \\
1987 & 0.132957 &       &       & \citet{Udalski88}            \\
1988 & 0.132816 &       &       & \citet{Andronov92}           \\
1988 & 0.132953 &       &       & \citet{Tremko92}             \\
1988 & 0.13306  & 0.076 & $B$   & this work                    \\
1990 & 0.13383  & 0.059 & $B$   & this work                    \\
1991 & 0.13386  & 0.055 & $B$   & this work                    \\
1992 & 0.13215  & 0.061 & $B$   & this work                    \\
1994 & 0.133160 & 0.102 & $B$   & \citet{Andronov99}           \\
1994 & 0.1323   & 0.18$^a$ & $B$ & \citet{Belova13}            \\
1995 & 0.1338   & 0.13$^b$& $B$ & \citet{Belova13}           \\
1995 & 0.13369  & 0.108 & $B$   & \citet{Kraicheva99}          \\
1996 & 0.13424  & 0.082 & $B$   & \citet{Kraicheva99}          \\
{\it 1997} & {\it 0.14926}  & {\it 0.13$^c$} & $V$ & \citet{Skillman98}\\
{\it 1997} & {\it 0.14931}  & {\it 0.150} & $V$  & this work           \\
{\it 1997} & {\it 0.14923}  & {\it 0.136} & $B$  & \citet{Kraicheva99} \\
{\it 1998} & {\it 0.14961}  & {\it 0.130} & $B$  & \citet{Kraicheva99} \\
{\it 1999} & {\it 0.14890}  &             &      & \citet{Wu02}        \\
{\it 2001} & {\it 0.14844}  & {\it 0.169} & $V$ & this work            \\
{\it 2001} & {\it 0.1500 }  & {\it 0.15$^d$} & $V$ & \citet{Belova13}\\
{\it 2004} & {\it 0.14860}  & {\it 0.136} & $V$ & this work         \\
{\it 2004} & {\it 0.1483 }  & {\it 0.20$^d$} & $R$ & \citet{Belova13}\\
2005 & 0.132322 &       &     & \citet{Kim09}                \\
2005 & 0.132624 & 0.094 & $V$ & this work                    \\
2007 & 0.1324   &       &     & \citet{Weingrill09}          \\
2007 & 0.13302  & 0.102 & $V$ & this work                    \\
2012 & 0.132874 & 0.116 & $V$ & this work                    \\
2014 & 0.132799 & 0.152 & $V$ & this work                    \\
2017 & 0.132643 & 0.154 & $V$ & this work                    \\
\hline
\multicolumn{5}{l}{$^a$ estimated from fig.~3 of \citet{Belova13}} \\
\multicolumn{5}{l}{$^b$ estimated from fig.~4 of \citet{Belova13}} \\
\multicolumn{5}{l}{$^d$ estimated from fig.~1 of \citet{Skillman98}} \\
\multicolumn{5}{l}{$^d$ estimated from fig.~2 of \citet{Belova13}} \\
\end{tabular} \\
\end{table}

The precession period of a tilted disk depends on the mass ratio of the
stellar components, the tilt angle and the mass distribution in the disk.
In the approximation of \citet{Larwood98} the latter resumes to the
disk radius. Keeping the tilt angle and, of course, the mass ratio fixed,
eq.~(4) of \citet{Larwood98} shows that, in order to explain the observed
scatter of the superhump period, changes of the disk radius of the
order of 10 percent (or less, considering that a significant part of the
scatter may be due to uncertainties to measure the period in small data
sets) are required. 

Whenever the corresponding information is available, the full amplitude of
the modulation is also listed in 
table~\ref{long term period stability}\footnote{Some references quote 
the half amplitude. In these cases the respective values were doubled.}
together with the photometric band to which it refers. The negative superhump
amplitude is quite variable with no clear systematic trends. Particularly 
striking is the huge amplitude observed in 1979. The 1992 data, consisting of 
five $UBV$ light curves and four light curves observed in the Stiening system,
permit an assessment of the colours of the superhump. Neglecting the slight
differences of the effective wavelengths of the passbands in the two systems, 
I find an average amplitude
of $\Delta U = 0.31$, $\Delta B = 0.35$, $\Delta V = 0.42$ and 
$\Delta R = 0.53$. Thus, it increases to the red, suggesting that the
modulation arises mainly in the outer, cooler parts of the accretion disk.

\subsection{QPOs and flickering}
\label{QPOs and flickering}

\citet{Smak69} were the first to draw attention to QPOs 
in TT~Ari occurring on a time-scale of 14 -- 20 min. 
They are readily visible in many light curves as flares with an amplitude
of the order of 0.2~mag (see, e.g., the lower frame of
Fig.~\ref{overall-lc}), although they often cannot unequivocally be 
separated from random flickering.

Much effort has been invested in the past to characterize these QPOs and
to find some regularity in their behaviour. While some authors content
themselves to quote some favoured periods seen in their light curves
\citep[e.g.,][]{Williams66, Udalski88, Hollander92}, 
others searched for systematic features in their long term behaviour 
\citep{Kraicheva97, Kim09}. \citet{Semeniuk87}
claim the detection of a systematic decrease of the QPO period
from 27~m in 1961 to 17~m in 1985. This is contested by \citet{Tremko96} 
who instead of a systematic trend suspect the presence of several 
preferred QPO frequencies. This view is also supported by \citet{Andronov99}
who find that the power spectra of their data contain significant 
peaks in the wide range between 24 and 139 d$^{-1}$ with the highest of them
corresponding to periods of 21 and 30~m. \citet{Kraicheva99} also
found the QPOs to be highly unstable with a coherency limited  to 3 -- 8 
cycles and power spectrum peaks detected in the 40 -- 120 d$^{-1}$ range. 
From an autoregressive analysis they suggest that the QPOs are most
probably generated by some stochastic process such as flickering. This
blurs the border line between the two phenomena \citep[see also][]{Bruch14}.

\subsubsection{QPO frequencies and their evolution}
\label{QPO frequencies and their evolution}

\begin{figure}
	\includegraphics[width=\columnwidth]{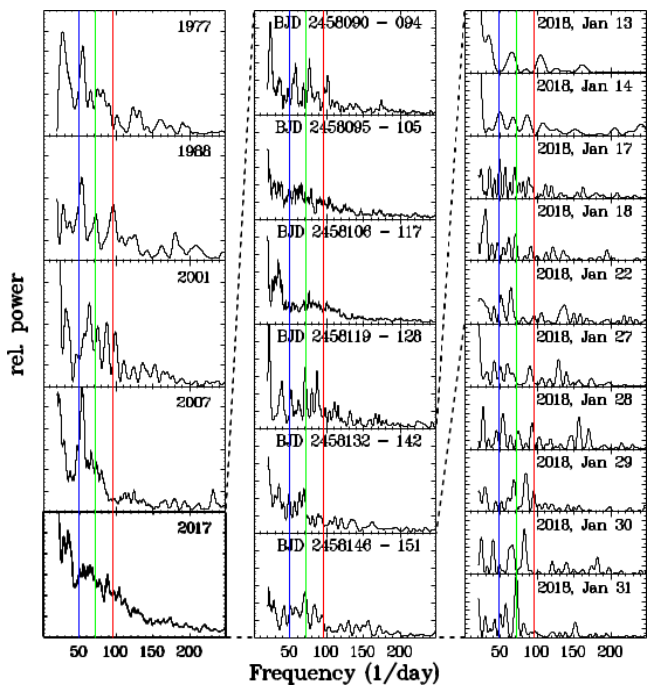}
      \caption[]{Examples of power spectra of light curves sampled on different 
                 time-scales: combined seasonal light curves (left), combined
                 light curves of intervals encompassing several days (middle),
                 and light curves of individual nights. The frequency range 
                 is restricted to the domain occupied by QPOs. The coloured
                 vertical lines have the sole purpose to guide the eye and
                 correspond to periods of 15 (red, right), 20 (green, middle) 
                 and 30 (blue, left) minutes.}
\label{qpo-1}
\end{figure}

The wealth of the present data confirms the notion that not much regularity
exists in the occurrence of the QPOs, except for a preference for a broad 
but limited frequency range. This is exemplified in Fig.~\ref{qpo-1}
which contains the power spectra of a selection of light curves on different
time-scales. For the left hand column, combined seasonal light curves of 5
different years were used, with intervals of roughly a decade between them
(noting that the number of contributing individual light curves differs 
greatly).
The central column contains the power spectra of the different time intervals
in 2017, as defined in Table~\ref{period stability}. Finally, the right hand
column is based on the light curves of the individual nights of the last two
time intervals. The coloured vertical lines, corresponding to periods of
15 (red, right), 20 (green, middle) and 30 (blue, left) minutes, 
are not meant to indicate specific
features in the power spectra but have the sole purpose to guide the eye
through the multitude of structures. The great variety of the power spectra
in the figure indicates the chaotic frequency behaviour of the QPOs.

\begin{figure}
	\includegraphics[width=\columnwidth]{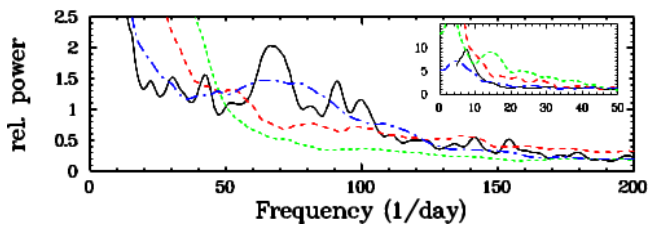}
      \caption[]{Normalized weighted average of the power spectra of
                 individual light curves of TT~Ari. The black graph
                 refers the high state, the coloured ones to sections 1
                 (red, dashed), 2 (green, dotted) and 3 (blue, dashed-dotted) 
                 of the 2009 -- 2011 low
                 state (see text for details). The insert shows the low
                 frequency part of the spectra on an expanded scale.}
\label{qpo-2}
\end{figure}

The overall frequency range occupied by the QPOs can be assessed from the
weighted average of the power spectra of all individual light curves, choosing
a weight proportional to the total time base encompassed by the nightly
data sets. The normalized sum of all power spectra, shown as a black graph
in Fig.~\ref{qpo-2}, 
has a dominant broad peak with its maximum corresponding to a period of 
21.4~m and a total width corresponding to a period range between 18
and 25~m. Both, at lower and at higher frequencies, fainter peaks appear. 
Without trying to assess their formal significance, they indicate that QPOs
appear to occur within some preferred period intervals between about 15
and 50~m, as claimed previously by \citet{Tremko96} and \citet{Andronov99}.

Assessing QPOs on the basis of power spectra of nightly light curves may be
misleading. Stacked power spectra permit a more detailed view of their
evolution over the time-scale of hours. For this purpose I selected some
long and high quality data sets: two unfiltered light curves taken in 1977 by 
E.\ Nather and the $B$ band of two Stiening light curves observed in 1992 
by E.\ Robinson. First, a filtered version, using a Savitzky-Golay filter 
\citep{Savitzky64}, was subtracted from the original data, removing all 
modulations on time-scales above about 90 min (thus, the superhumps). Then,
Lomb-Scargle periodograms for sections of a data train, 0.04~d long,
were constructed, allowing for a strong overlap of 0.039~d between subsequent 
sections. Thus, each segment covers approximately 3 cycles of a typical
QPO. The individual power spectra were then stacked on top of each 
other, resulting in a 2D representation (frequency vs.\ time)
\citep[for a more detailed description of this technique, see][]{Bruch14}. 

\begin{figure}
	\includegraphics[width=\columnwidth]{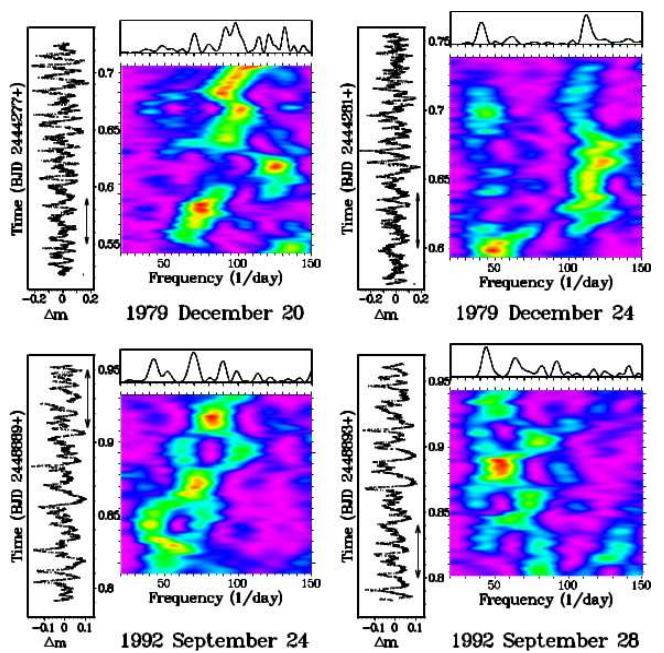}
      \caption[]{Stacked power spectra of TT~Ari in four nights. The
                 images show the power (colour coded) as a function
                 of frequency (horizontal axis) and time (vertical axis).
                 On the left, the light curves are shown after subtraction
                 of variations on time-scales of more than 90~min. The double
                 arrow indicates the length of the individual sections used
                 to calculated the stacked power spectra. Thus vertical
                 structures in the 2D images extending over
                 less than the length of the arrow are not independent.
                 On top of each stacked power spectrum the 'conventional'
                 periodogram of the entire data set it drawn.}
\label{stacked-ps}
\end{figure}

The results are shown in Fig.~\ref{stacked-ps} which, for each of the four
nights, shows the high-pass filtered light curve in the left frames,
the 'conventional' power spectrum (using the entire light curve) on top,
and the stacked power spectrum as a 2D colour coded image.
The double arrow drawn beneath the light curves indicates the length of
the individual data segments used to calculate the power spectra. Thus,
any vertical structures in the images smaller than this length are not
independent from each other.

The stacked power spectra suggest patterns in the occurrence of the QPOs.
On 1977 December 20 (top left), and 1992 September 24 and 28 (bottom) there 
appears to be
an evolution of the QPO frequency to higher or lower values over the
time interval covered by the light curves. On 1977 December 20, this is
interrupted by the short appearance of a signal at higher frequencies but is 
then resumed, while on 1992 September 24, the QPOs occasionally seem to split 
into two branches. On 1977 December 24 (top right), QPOs at two distinct 
frequencies are present, one getting fainter when the other grows stronger.
The 1992 September 24 results are a good example of a misleading 
'conventional' power spectrum. It suggests 
the presence of three distinct frequencies, while the stacked version rather
indicates an evolution of both, the frequency and the strength of the QPOs.
\citet{Semeniuk87} claim coherence of the QPOs over intervals of at
least 6~h (and possibly even from night to night). While the present
data, in most cases, cannot confirm coherence, they show that the same
train of events may persist over several hours, gradually changing
frequency.

\subsubsection{Flickering and QPO amplitudes}
\label{Flickering and QPO amplitudes}

The definition of the amplitude (or strength) of the ubiquitous flickering
in the light curves of cataclysmic variables is not trivial and, to my
knowledge, has never been done rigorously and objectively. The standard
deviation, variance or rms-scatter of individual magnitude measurements 
in a light curve is occasionally taken to be a measure of the 
flickering strength \citep[e.g.,][]{Dobrzycka96, Dobrotka15}.
This is certainly to be preferred to simply 
adopting the magnitude difference between the peaks of flickering flares 
and the troughs between them, since the latter depends more strongly on 
individual and random features in a light curve, while the former take 
into account their average. Even so, there are some pitfalls which a more 
rigorous approach has to consider.

\begin{figure}
	\includegraphics[width=\columnwidth]{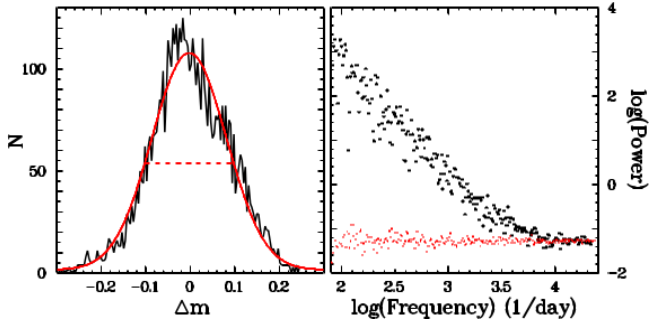}
      \caption[]{{\it Left:} Illustration for the measurement of the
                 flickering strength: Histogram of the distribution of
                 data points in a flickering light curve (black) and 
                 best fit Gauss curve (solid red). Its FWHM (broken red line)
                 is defined as the flickering amplitude.
                 {\it Right:} Illustration for the determination of the
                 noise level in a flickering light curve: Black dots
                 are a log-log representation of its power spectrum. The
                 smaller red dots represent the average power spectra of 10
                 light curves of pure Gaussian
                 noise, sampled in the same way as the real light curve,
                 with the noise amplitude adjusted to the flat high
                 frequency part of the power spectrum of the real light curve.}
\label{fl-tech}
\end{figure}

Postponing a more detailed description to an upcoming paper (Bruch, in
preparation) I hereafter briefly summarize the method used to quantify the 
strength of the flickering in CVs. Most importantly, variations unrelated
to flickering have to be removed. If these have longer time-scales (e.g.,
orbital modulations or superhumps) this is easily done by subtracting a 
filtered version of the original data. I use a Savitzky-Golay filter 
with a cut-off time-scale of 60 minutes. In the case
of TT~Ari this does not separate the QPOs from flickering \citep[if such a
separation is meaningful at all; see the brief discussion in][]{Bruch14}.
Thus, here I will study the combined strength of QPOs and flickering which
is in most light curve dominated by the QPOs. Variations on short time
scales, distinct from flickering 
\citep[such as dwarf nova oscillation, see][]{Warner04} 
are only resolved in high time resolution light curves and
generally of low amplitude. They can therefore be neglected.
A histogram is constructed from the individual magnitude data in the 
high-pass filtered light curve. As is shown in the
left frame of Fig.~\ref{fl-tech}, it can in general well be
approximated by a Gaussian. I take the full width at half maximum (FWHM) as
a proxy of the flickering strength and denote it as $A_{60}$, considering
that it only contemplates variations on time-scales smaller than 60~min. 
Note that $A_{60}$ is smaller than but proportional to the average amplitude 
of flickering flares. It enables the comparison of the
flickering strength in different light curves and in different systems.

$A_{60}$ is a measure similar to the variance of data points in a light
curve. However, some corrections are required. The most obvious one is
a correction for noise in the data which widens the distribution of data
points. If the noise can be regarded as Gaussian, the variance of the 
observed data is simply the sum of the variance due to noise and to
real variations. A correction is then easy. The problem obviously lies in
the determination of the noise level in the data. Here, the well known
red noise characteristic of flickering helps. As shown in the right
frame of Fig.~\ref{fl-tech}, the power spectrum of a flickering light 
curve (larger black dots), drawn on a double-logarithmic scale, slopes
downward linearly at high frequencies, but levels off to a constant at
very high frequencies, where it is dominated by white noise. Generating
a power spectrum of pure Gaussian noise (smaller red dots in the figure), 
sampled in the same way as the real light curve, with an amplitude such that it 
matches the flat part of the flickering light curve power spectrum, provides
the noise level in the data\footnote{Note that the power spectra must not
be normalized. Therefore the Lomb-Scargle algorithm is not adequate. 
Instead, I use the algorithm suggested by \citet{Deeming75} for this purpose.}.

Another correction concerns the time resolution of the data. Larger
integration times act as a filter reducing the height and depth of
extrema. Thus, a reduction of $A_{60}$ to a reference time resolution,
which I define to be 5~s, is required. This effect depends on the time-scale 
on which flickering events predominantly occur, which may differ from one 
object to the other. It can be accounted for, determining a correction factor
as a function of the time resolution. For this purpose, a high quality
light curve observed with the reference (or higher) time resolution is
binned into intervals, simulating a reduced resolution. Comparing $A_{60}$
measured at the reference and at other time-scales provides the correction
factor.

A third correction, required in some cases, can be omitted here. 
It concerns the contribution of the secondary star light that dilutes the
flickering. However, in TT~Ari in the high state it is wholly negligible. 

\begin{figure}
	\includegraphics[width=\columnwidth]{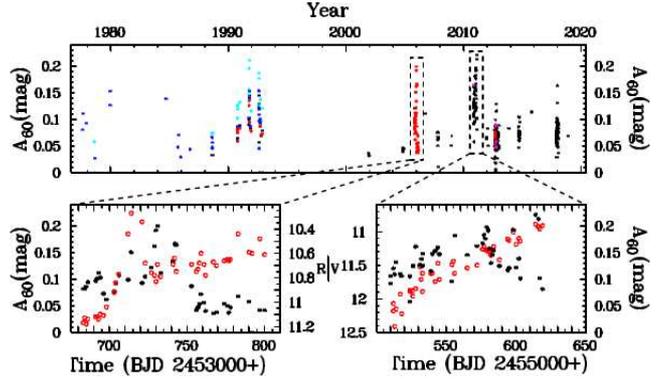}
      \caption[]{{\it Top:} Flickering amplitude $A_{60}$ as a function of 
                 time for observations taken in $U$ (cyan), $B$ or white 
                 light (blue), $V$ {black}, $R$ (red) and $I$ (magenta).
                 {\it Bottom:} Amplified view of two limited time intervals
                 in 2005 and 2010 -- 2011 marked in the
                 upper frame. In addition to $A_{60}$ (black dots)
                 the mean nightly magnitudes are shown as red circles.}
\label{fl-amp}
\end{figure}

$A_{60}$ was measured in all high state light curves of TT~Ari and is shown 
as a function of time 
in the upper frame of Fig.~\ref{fl-amp}. The colours reflect the observed 
passbands: $U$ (cyan), $B$ or white light\footnote{All white light observations
used here were taken with photomultipiers which have a blue response. Together
with the spectral energy distribution of CVs this leads to an isophotal
wavelength similar to that of the $B$ band.} (blue), $V$ (black), $R$ (red) and 
$I$ (magenta). $A_{60}$ scatters significantly. This is partly due to 
measurement errors caused mainly by two effects: (i) in light curves with a 
comparatively small number of data points the histogram of their distribution 
is not well defined, leading to uncertainties of the FWHM of the fitted 
Gaussian; and (ii) in light curves with a coarse time resolution the 
Nyquist frequency 
is such that the useful part of the power spectrum does not reach the plateau
at high frequencies, enabling only to determine an upper limit to the noise
level. However, these effects cannot mask real and systematic variations of 
the flickering amplitude which are particularly evident in two sections of 
the long term light curve.

These sections are shown in more detail in the lower frames of 
Fig.~\ref{fl-amp}. Here, the red circles indicate the
average magnitude of the respective light curves. Both sections correspond
to periods when TT~Ari underwent particular phases in its long term 
behaviour. The left frame refers to the data of \citet{Kim09}. As
was mentioned in Sect.~\ref{Re-analysis of the Kim et al. (2009) data},
TT~Ari recovered from a minor excursion to fainter magnitudes. $A_{60}$ 
assumed an intermediate level when the system was still at the bottom
of this excursion, goes through the maximum and then settles at a lower
level when TT~Ari assumed its long-term high state magnitude after a
short term maximum. It is interesting to note that the maximum of $A_{60}$
is assumed only about two weeks after the maximum in brightness. Thus,
there is no strict correlation between flickering amplitude and system
magnitude. This is also seen in the second section (right lower frame of
Fig.~\ref{fl-amp}) which corresponds to the very end of the 
recovery of TT~Ari from the low state in 2009 -- 2011. Here,
$A_{60}$ rises from an intermediate level, goes through a maximum, 
and then declines to the same level as before, while the brightness
of the system rises steadily during the entire time interval.

\begin{figure}
	\includegraphics[width=\columnwidth]{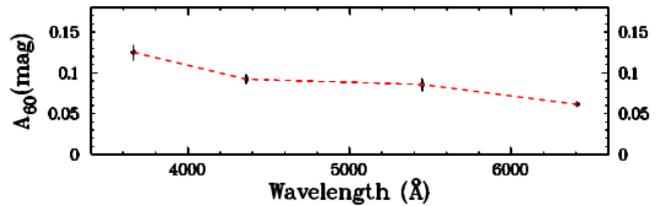}
      \caption[]{Wavelength dependence of the flickering amplitude in
                 TT~Ari.}
\label{fl-spectrum}
\end{figure}

Concluding this section, I briefly investigate the wavelength dependence of
the flickering amplitude. For this purpose I restrict myself to the 
simultaneous multicolour observations provided by A.\ Hollander (Walraven 
system), T.\ Schimpke (Johnson $UBV$) and E.\ Robinson (Stiening 
$UBVR$ system) in order not to mix data of different passbands obtained 
at different epochs. The
slight differences of the effective wavelengths of the passbands in
the three photometric systems are neglected, adopting the effective 
wavelengths of the $UBVRI$ system as tabulated by \citet{Bessell05}. 
The average value of $A_{60}$ in the $UBVR$ passbands is shown as a function 
of wavelength in Fig.~\ref{fl-spectrum} which reflects the well known 
flickering property in CVs to increase in strength to the blue and UV.
The error bars are mean errors of the mean which are a better representation
for the accuracy of the average value than the standard deviation. It must 
be noted, however, that the figure does not reflect directly the spectral 
energy distribution
(SED) of the flickering light source. This would only be true if the constant
(i.e., not flickering) components of the system (basically the quiet light
of the accretion disk), upon which flickering is superposed, had the same
SED. This is not necessarily the case.

\section{The 2009 -- 2011 low state}
\label{The 2009 -- 2011 low state}

In contrast to the high state, the two deep low states observed in TT~Ari
in 1979 --- 1985 and 2009 -- 2011 have received much less attention. 
The first one was treated in some detail by \citet{Shafter85} and 
later by \citet{Gaensicke99}, while \citet{Hutchings85}
presented some complementary observations. To my knowledge, the only 
paper discussing observations of the 2009 -- 2011 low state was
published by \citet{Melikian10}. Apart from some time resolved light
curves observed by \citet{Shafter85} not much detailed photometry
performed in this state has been studied. In particular, the AAVSO data
of the first low state consist only of isolated visual observations which
are of little use for the investigation of variations on short time-scales.
This is different when it comes to the second low state, parts of which have
extensively been covered with high time resolution light curves obtained by
AAVSO observers.

\begin{figure}
	\includegraphics[width=\columnwidth]{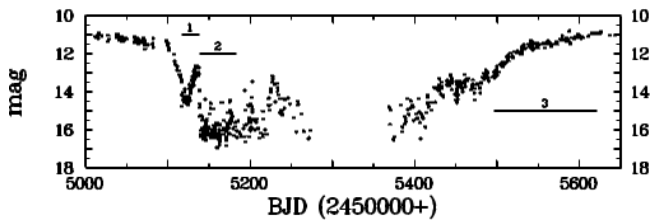}
      \caption[]{Light curve of TT~Ari during its 2009 -- 2011 low state,
                 binned into 1~day intervals. The
                 horizontal lines indicate 3 sections discussed
                 in more detail in the text.}
\label{low-state-lc}
\end{figure}

An expanded view of the long-term light curve during the 2009 -- 2011 low
state is shown in Fig.~\ref{low-state-lc}, where the data have been binned
in 1~day intervals. In view of the strong variations in comparison to the
colour indices of TT~Ari, for visualization purposes measurements in different
passbands are not distinguished in the figure. The overall evolution is such 
that an initial drop by 3.3~mag, lasting 26~d, is followed by a $\approx$8~d
plateau and a subsequent steady rise by 1.7~mag over $\approx$12~d. 
This is reminiscent of the start of the 1979
-- 1985 low state (Fig.~\ref{overall-lc}) when TT~Ari also recovered almost
to the high state magnitude after an initial drop, albeit on longer time 
scales. This first recovery is followd by a precipitously drop by
3.3~mag to $\approx$16.3~mag within less than 3~d. After
a plateau lasting about 80~d (much of the scatter in this phase is caused
by the specific sampling of strong intra-night variations; see below)
a rapid brightness increase to another local maximum 
and a subsequent drop to the same low level is observed. Following
the seasonal gap in the observations, TT~Ari starts a continuous recovery from
the low state, exhibiting four phases with distinct gradients in
the light curve.

Three sections of the low state light curve are of particular interests. These
are marked by horizontal lines in Fig.~\ref{low-state-lc}, and will
subsequently be discussed in more detail.

\begin{figure}
	\includegraphics[width=\columnwidth]{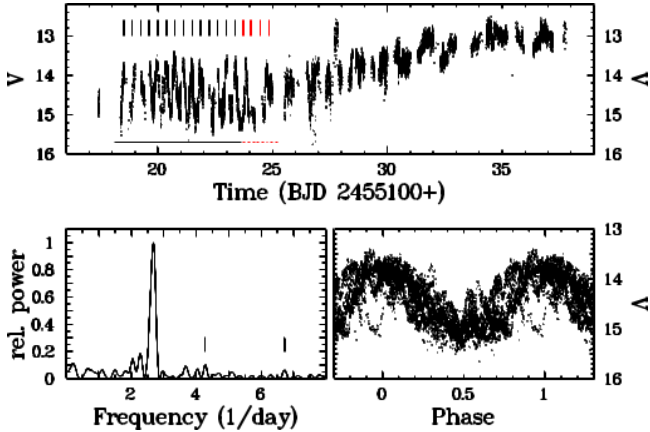}
      \caption[]{$V$ band light curve of the plateau phase and the subsequent
                 rise after the initial drop of TT~Ari into the low state 
                 (upper frame). The
                 horizontal line below the light curve indicates a section with
                 strong oscillations which are coherent at first (black) but 
                 later loose coherence (dotted red). The vertical tick marks 
                 are the predicted times of maxima calculated using the 
                 period corresponding to the peak frequency of the power
                 spectrum (lower left frame; the two marked features are
                 discussed in the text) based on the part of the
                 light which exhibits coherent modulations. The lower right
                 frame contains the light curve folded on the period.}
\label{low-state-detail-1}
\end{figure}

\subsection{Coherent large amplitude modulations}
\label{Coherent large amplitude modulations}

The first section
refers to the short lived plateau and subsequent rise after the initial
plunge from the high state. The upper frame of Fig.~\ref{low-state-detail-1}
presents an expanded view of this section, maintaining the original time 
resolution. Apparently periodic large amplitude variations are immediately
evident during the plateau phase. It turns out that they lose coherence 
towards the end. Therefore two intervals of the plateau phase were defined, 
distinguished by the black and dotted red horizontal lines below the light 
curve. The power spectrum of the first of these intervals is shown in the 
left lower frame of Fig.~\ref{low-state-detail-1}. The outstanding peak 
corresponds to a period of $0.372 \pm 0.011$~d. The vertical bars in the 
upper frame of the figure, spaced with this period, indicate the excellent 
alignment with the light curve maxima in the first interval, which is lost 
during the latter part of the plateau phase. The lower right frame shows 
the coherent part of the light curve, folded on the period.
The total amplitude of the variations amounts to $\approx$1.2~mag.

The power spectrum contains many minor peaks that, however, are all
statistically significant. For most of them no simple relation 
between their frequencies is evident. 
It is worthwhile, however, to draw attention to a small peak \citep[but still 
with a maximum power more than 10 times stronger than the 0.0001 false
alarm probability level as calculated using the prescription of][]{Bruch16}
close to 6.7 d$^{-1}$ (marked with a vertical bar in the figure),
corresponding to a period of $0.1487 \pm 0.0014$~d. Coincidence or not,
this is very close to the centre of the period range of the positive
superhumps observed in 1997 -- 2004. Moreover, curiously, this period and
the main period of 0.372~d, within less than 1 percent of their formal errors,
are multiples (2$\times$ and 5$\times$, respectively) of a basic period
of 0.0744 days. They may therefore be regarded as overtones of each other.
Another peak (also marked in the figure) corresponds to a period of 
$0.233 \pm 0.003$~d, close to (but beyond the formal error limit) three
times this basic period. I leave it open if the relationship of these
periods with the positive superhump period has any physical meaning, or if 
I am over interpreting the data.
  
After the plateau phase a steady increase of the brightness sets in. While
intra night variation on the level of some tenths of a magnitude still occur,
no evidence of any regularity from night to night could be detected. Flickering
is present during the entire section discussed here.

\begin{figure}
	\includegraphics[width=\columnwidth]{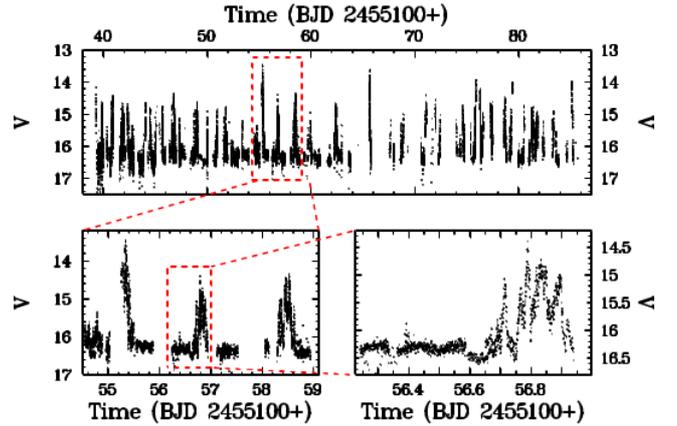}
      \caption[]{Light curve during the deep low state
                 of TT~Ari together with views with increasing
                 time resolution of a light curve section.}
\label{low-state-detail-2}
\end{figure}

\subsection{The deep low state}
\label{The deep low state}

The second section refers to the bottom of the low state (hereafter: the
deep low state) after the second rapid drop in brightness. The upper frame
of Fig.~\ref{low-state-detail-2} 
contains the light curve of the entire section, while a small
part of it is shown in increasing detail in the lower frames. Disregarding some
excursions to even lower magnitudes (which may well be due to measurement
uncertainties at this quite low brightness level for the instruments used
by many AAVSO observers) the light curve shows a quite well defined lower
limit of about 16.3~mag. 

Using observations taken
during the 1979 -- 1985 low state, \citet{Gaensicke99} estimate the visual
magnitude of the secondary star to be 19.07~mag and the temperature
of the white dwarf as 39\,000~K. The latter is a close match to EC~11437-3124
which, at a temperature of 38\,810~K \citep{Gianninas11}, has $V = 17.32$.
At 208.6~pc \citep{Bailer-Jones18} its distance is slightly less than
that of TT~Ari. Correcting for this difference it should have $V = 17.53$ at
the distance of TT~Ari. This is an upper limit for the brightness of the
TT~Ari white dwarf because \citet{Gianninas11} estimate a mass of 
$0.69\ M_\odot$ for EC~11437-3124, less than the (uncertain) mass of TT~Ari
that should thus have a smaller radius and be less luminous. The combined 
light of the stellar components of TT~Ari should therefore not be brighter
than $V = 17.3$, a magnitude fainter than the observed lower limit in the light
curve. This implies (an) additional light source(s) in the system, which may
be a residual accretion disk.

\subsubsection{Intermittent activity}
\label{Intermittent activity}

The lower brightness limit of 16.3~mag is defined by quiet phases with are 
interrupted by strong flares with amplitudes typically of $\approx$1.5~mag but 
that can reach almost 3~mag. There is no strict periodicity in the 
occurrence of the latter, but the median
interval of 1~d between them is quite well defined, and they last for
about 5 -- 8~h. As is best seen in the lower right frame of 
Fig.~\ref{low-state-detail-2}, individual flares are heavily structured. 
\citet{Melikian10} have already drawn attention to this flaring
activity of TT~Ari during the low state. However, their limited amount of
data did not permit a more thorough characterization of this behaviour.

The quiet phases are seen only during a limited time interval, covering most
of the minimum between the first and the second re-brightening during the
low state. The sampling of the light curve becomes less dense during the 
second half of the section shown in Fig.~\ref{low-state-detail-2}, but it 
appears that the number and duration of the quiet phases decreases. At 
later times (not show in detail), the sampling becomes even
less complete which makes any statement about the occurrence of quiet phases
less secure. However the last time resolved light curve without significant
variability was observed on 2009 December 25, i.e., before the second
re-brightening. All subsequent observations contain strong irregular 
variations on the time-scale of hours.

The lack of observations makes it difficult to assess if the alternation
of quiet phases and strong flaring activity also occurred during the first
deep low state of TT~Ari. The AAVSO long term light curve, restricted to
visual magnitude estimates, does not have sufficient time resolution to 
answer this question. But it contains some points which lie about 1~mag
above the mean low state level, suggesting flares. Moreover, the limited
time resolved observations discussed by \citet{Shafter85} includes
light curves at a similar brightness level as the present quiet phases with
only low-scale flickering over time intervals of $\approx$3~h. Thus, the
overall behaviour may have been similar during the 1979 -- 1985 low state.

The alternation between quiet and flaring phases during the low state of
VY~Scl stars is not unheard of. While in most such systems the available
low state observations are not sufficient to explore this question, the
long uninterrupted high cadence observations of MV~Lyr by the {\it Kepler} 
satellite enabled \citet{Scaringi17} to detect a similar behaviour
in that star. The variations seen in the bottom frame of their fig.~1 bear 
much resemblance with the TT~Ari deep low state behaviour, although some 
differences are also obvious. Even not being strictly periodic, the flares 
occur much more regularly in MV~Lyr. The burst duration and typical intervals
are 30~m and 2~h, respectively, versus 5 -- 8~h and 1~d in TT~Ari. These 
numbers make the duty cycle of the flares quite similar in both systems.
The flare amplitudes are also similar, being typically $\approx$1.5~mag 
and occasionally even larger. 

\citet{Scaringi17} explain the flaring behaviour of MV~Lyr within a 
magnetic gating model, where a magnetic field of the primary star disrupts
the inner disk out to the co-rotation radius, forming a centrifugal barrier.
Matter in the disk then accumulates outside the barrier, increasing the 
pressure until it is overcome, permitting accretion onto the white 
dwarf and releasing bursts in the process. \citet{Scaringi17} show that, 
depending mainly on the mass accretion rate and the white dwarf spin period, 
a magnetic field as low as
22 kG, too small to be directly detectable, may be sufficient to create the
barrier. Once the accretion rate rises above a certain limit and thus the
system becomes brighter, the magnetic field is no more able to create the
barrier and the intermittent flaring behaviour stops.
 
A similar model may be able to explain the deep low state behaviour of
TT~Ari. Not knowing important system parameters, in particular the rotation
period of the white dwarf which determines the co-rotation radius, it is
difficult to assess if the model is compatible with the details of the
observations along the lines investigated for MV~Lyr by \citet{Scaringi17}. 
An argument in favour of the presence of a weak magnetic field is the
finding of \citet{Thomas15} in their SHP simulations that a field in the 
kilogauss regime can lead to the emergence of negative superhumps in novalike
systems. On the other hand, if matter accumulates outside the centrifugal 
barrier between flares, a gradual increase of the system brightness during the
non-flaring phases may be expected but is not seen. This is equivalent to
the lack of brightening, contrary to theoretical predictions, of dwarf
novae between outbursts \citep{Lasota01}.

A further drawback of the magnetic gating model is the fact that, while 
explaining the alternation between quiet and flaring phases during the deep 
low state, another explanation is required for the continuous strong and 
either a-periodic or periodic (see 
Sect.~\ref{Coherent large amplitude modulations}) variations seen at
higher brightness levels throughout the low state. Assuming as a single 
mechanism modulations of residual mass transfer from 
the secondary star may be able to account for both. The involved time-scales
of hours are easily compatible with dynamical (free fall) time-scales in the 
binary system.

While it is a broadly accepted view that the low states in VY~Scl stars 
(as well as those observed in some members of the magnetic subclass of CVs) 
are caused by a decrease or cessation of mass transfer from the secondary 
components, the reason for this decrease are less clear. The most thoroughly
discussed scenario, first elaborated by \citet{Livio94}, is that of star spots 
crossing the L$_1$ region. The cooler gas in the spots has a smaller scale
height than the undisturbed stellar atmosphere. It can therefore detach from
the L$_1$ point, reducing or inhibiting mass transfer. Elaborating this 
scenario further and applying it to the low state of AM~Her, \citet{Hessman00}
assume not a single spot but a heavily structure star spot region to be 
responsible for strong variations of the mass transfer rate in AM~Her 
and consequently of the brighness during this state. A similar 
situation may
explain the low state variations occurring over the course of weeks in
TT~Ari. The strong modulations on the time-scale of hours may then be due 
to hydromagnetic exchange instabilities, as already
speculated by \citet{Livio94}. This leaves open only the reason for the 
strict periodicity seen in the first part of section 1.

\begin{figure}
	\includegraphics[width=\columnwidth]{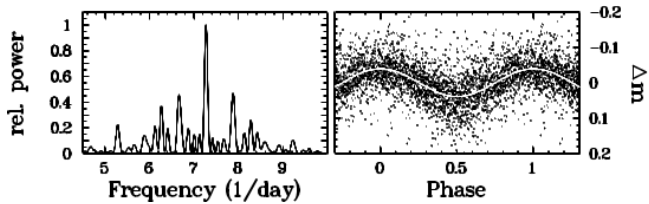}
      \caption[]{{\it Left:} Power spectrum of non-flaring phases of the
                 light curve of TT~Ari during the deep low state. The
                 dominant peak occurs at the orbital frequency.
                 {\it Right:} Low state light curve folded on the orbital
                 period. The white line is a least squares sine fit.}
\label{low-state-orbital}
\end{figure}

\subsubsection{Orbital variations}
\label{Orbital variations}

During the quiet phases the brightness of TT~Ari is not quite constant. In 
spite of the significant noise at this low light level, some light curves
clearly exhibit variability. It is different, however, from the
flickering activity observed by \citet{Shafter85} at a similar 
magnitude and noise level. While flickering may still be present in the
data discussed here, but may be more difficult to detect above the noise
because of a much lower time resolution than that of the \citet{Shafter85}
data, more obvious are
consistent variations on the time-scale of hours. I selected eight light 
curves in the time interval between 2009 November 17 and 25, where 
these modulations are best seen. After subtracting 
night-to-night variations they were subjected to the Lomb-Scargle algorithm.
The resulting power spectrum is shown in the left frame of 
Fig.~\ref{low-state-orbital} and is dominated by a strong peak at 
7.270~d$^{-1}$. The secondary peaks are either 1~d$^{-1}$ aliases or can be 
explained as being caused by the window function. 

The peak frequency
corresponds exactly to the spectroscopic orbital period of TT~Ari. Thus,
for the first time, a clear photometric manifestation of the binary 
revolution is observed. The light curve, folded on the orbital period,
is shown in the right frame of Fig.~\ref{low-state-orbital}. A formal
sine fit (white curve) yields a total amplitude of 0.078~mag for
the modulation. The epoch of the maximum is BJD~2455152.716. The
spectroscopically determined epoch of the inferior conjunction of the
primary component, closest to the epoch of the present observations, cited
in table~3 of \citet{Wu02}, together with their orbital period,
yields a phase difference of $0.06 \pm 0.03$ between conjunction and light
curve maximum. Here, the error includes only the formal period uncertainty.
The true error may be larger. Thus, the modulations are consistent with
an illumination of the secondary star by the white dwarf, this extra light
being best visible to the observer at the inferior conjunction of the latter.

\begin{figure}
	\includegraphics[width=\columnwidth]{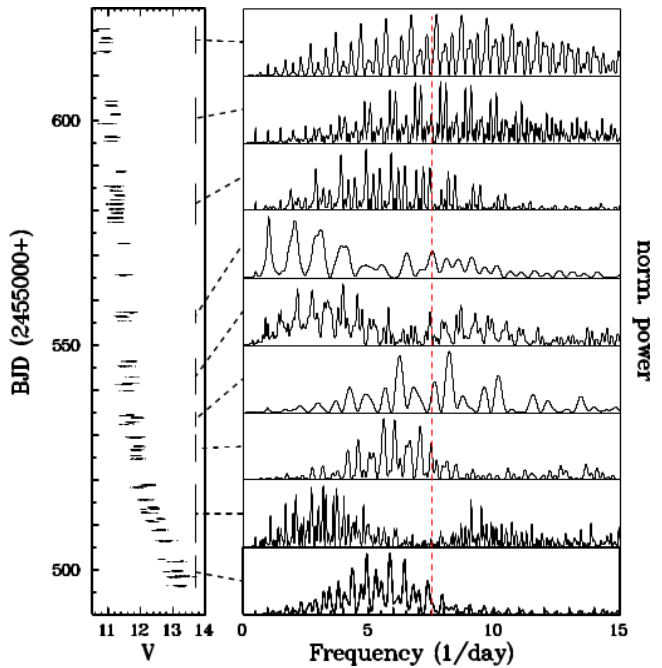}
      \caption[]{Light curve during the rise from the low to the high state
                 (left) together with power spectra (right) calculated 
                 from the light curve sections marked by vertical bars. The red
                 vertical line indicates the average frequency of the superhump 
                 modulation, derived from 
                 Table~\ref{long term period stability}.}
\label{rise}
\end{figure}

\subsection{The rise towards the high state}
\label{The rise towards the high state}

The third section refers to the rise from the low to the high state. The 
entire rise phase, starting even before section 3, can clearly be separated 
into four parts: a gradual rise followed by a plateau; thereafter a steeper
rise with a gradient of $0.034\, {\rm mag\, d{^{-1}}}$ sets in, before TT~Ari
approaches the high state at a much slower pace of 
$0.009\, {\rm mag\, d{^{-1}}}$.

The steep part and then the much flatter final rise are reminiscent of the 
observations of \citet{Honeycutt04} who
found distinct gradients during the start and/or the end of low states in
several other VY~Scl stars, with the steeper slope always occurring when
the system was fainter. Within the star spot scenario for the low states
\citet{Honeycutt04} interpret the different gradients as being caused by the
passage of the umbra (steeper slope) and penumbra (flatter slope) underneath
the L$_1$ point. A similar situation may be realized in TT~Ari. However, the
behaviour at the onset of the rise points at
a more complex situation (see also Sect.~\ref{Intermittent activity}).

Section 3 was also selected because it
is of interest in order to investigate if and, in case, when and at 
which brightness level TT~Ari develops the superhumps seen at later epochs.
For this purpose, the light curves were subjected to the same analysis which 
has been applied to the Kim et al.\ (2009) data in 
Sect.~\ref{Re-analysis of the Kim et al. (2009) data}. The results are shown
in Fig.~\ref{rise} which is organized in the same way as 
Fig.~\ref{Kim powerspectra}. No consistent picture emerges. At most, the 
power spectra contain some indications that modulations on the time-scales
of hours repeat over a few nights. In particular, although in some of the
spectra a peak aligns approximately with the red vertical line which 
indicates the average superhump frequency calculated from 
Table~\ref{long term period stability}, these may well be chance alignments.

\subsection{Stochastic variations during the low state}
\label{Stochastic variations during the low state}

Except for the strong 0.372~d oscillations observed during the first part
of section 1 and the slight orbital modulations during the quiet phases of
the deep low state, no regularly repeating variations could be detected during
the low state. However, the average power spectra of the individual light 
curves within the three sections defined in Fig.~\ref{low-state-lc},
calculated in the same way as has been done for the high state light
curves (Sect.~\ref{QPO frequencies and their evolution}), exhibit 
systematic differences. They are shown in Fig.~\ref{qpo-2} as red (dashed), 
green (dotted)0 and blue (dashed-dotted) 
lines for sections 1, 2 and 3, respectively. During section 2 only
the flaring parts of the light curves were considered. Since due to the
smaller number of individual light curves the average power spectra are
rather noisy they have been smoothed in frequency by a bock filter of 
half-width 6~d$^{-1}$. The insert in the figure contains the low frequency
part of the power spectra (unsmoothed) on an expanded vertical scale.

The average power spectrum of section 1 (red) exhibits, after a steep decline 
from a strong maximum at low frequencies (a reflection of the 0.372~d 
oscillations), a steady decline towards higher frequencies. On a double
logarithmic scale it is strictly linear, as is the case for flickering
activity in CVs in general. During section 2 (green), when quiet intervals
alternate with strong flares, the high frequency
part also shows a behaviour typical of flickering, but there is a pronounced
peak (see insert in Fig.~\ref{qpo-2}) centred on 13.4~d$^{-1}$ ($\approx$1.8 h),
reflecting the time-scale of the main structures within flares. In both,
section 1 and 2, no indication for the occurrence of QPOs as observed
during the high state is seen. Finally, during section 3 (blue), 
while TT~Ari recovers from the low state, the average  
power spectrum contains a distinct hump between 40 and 90~d$^{-1}$ which
apparently indicates the re-appearance of QPOs. It is already present in 
the steeper part of the recovery from the
low state, but becomes stronger during the final rise.

\section{Conclusions}
\label{Conclusions}

Complementing previous studies, I have investigated variations in the 
novalike variable TT~Arietis with an emphasis on time-scale of hours and 
smaller, using an unprecedented wealth of data obtained over more then 40~yr 
during high and low states. During most of the time interval covered
by the observations, the system remained in its normal high state, suffering
only slight variations around its average magnitude. The main findings
during this stage are:

\begin{enumerate}
\item
The well known negative superhump, being replaced by a positive superhump
only during the limited time interval between 1997 and 2004, persists to
the present day. While within a given observing season its period was found
to be stable, small variations occur from year to year. The long term
average period is $0.13295 \pm 0.00067$~d (3.1908~h). The full amplitude
is quite variable, ranging from a staggering 0.33~mag in 1979 to low
values of about 0.06~mag in several other years. Confirming earlier reports
based on smaller data sets, in the two observing seasons, when the
quantity and quality of the present observations permit a respective
statement, a modulation on the precession period of the inclined accretion
disk is also seen.
\item 
In observations taken during the 2005 observing season, when TT~Ari 
recovered from a short lived excursion to a fainter magnitude and went
through a local maximum before settling down to the average high state
brightness, the superhump was initially absent but then rapidly emerged 
on the time-scale of only a few days.
\item
QPOs are present during the entire high state. They occur preferentially
in the quasi-period range between 18 and 25~m.
In some light curves the evolution of their frequency and amplitude could
be followed for several hours. In most of them a systematic evolution
towards higher or lower frequencies is seen, while in one night two
distinct frequencies appear to alternate.
\end{enumerate}

Of the two deep low states suffered by TT~Ari, that of 2009 -- 2011 was
very well covered by observations, permitting for the first time a
detailed characterization:

\begin{enumerate}
\item 
The overall low state light curve is strongly structured on the time
scale of tens of days. The rise to the high state occurs  in four stages
with distinct gradients.
\item
After the first plunge into the low state, but still a magnitude above
the deep low state attained later, during about five days TT~Ari undergoes
coherent large scale oscillations with an amplitude of $\approx$1.2~mag and
a period of 8.93~h. 
\item 
During the deep minimum at 16.3~mag TT~Ari alternates on time-scales of 
roughly one day between quiet phases and strong (1.5 -- 3~mag) flares.
These last for 5 -- 8~h and are highly structured on a preferred time-scale
of 1.8~h. During the quiet
phases the brightness is modulated on the orbital period. The phasing
of these variations is consistent with reflection of radiation of the
hot primary off the secondary star.
\item
Strong variations on the time-scale of hours are seen throughout the 
low state, except during the quiet phase at its deepest parts. They are
irregular but for the short coherent period at the beginning of the
low state mentioned earlier. No clear signal of a superhump is seen even 
during the final 
rise to the high state. However, the $\approx$20~m QPOs first re-emerge during
the steep part of the rise and grow stronger as the system approaches the
high state. 
\end{enumerate}

\section*{Acknowledgements}

This work is entirely based on archival data. I am grateful to all those
colleagues who have kindly put their data at my disposal, namely 
A.\ Hollander, E.\ Nather, R.\ Robinson, S.\ R\"o{\ss}iger, T.\ Schimpke 
and I.\ Semeniuk. In particular, I thank the numerous dedicated observers
of the AAVSO for their contributions, together with the AAVSO staff for 
maintaining the International 
Database. Without their efforts this work would not have been possible.
This research made use of the VizieR catalogue access tool, CDS, France
(DOI: 10.26093/cds/vizier).







\bsp	
\label{lastpage}
\end{document}